\documentclass[conference,letterpaper]{IEEEtran}
\IEEEoverridecommandlockouts
\usepackage{amsmath,amssymb,amsthm,bm,mathtools}
\usepackage{graphicx,booktabs,array,cite,xcolor}
\usepackage{tikz}
\allowdisplaybreaks[3]
\usetikzlibrary{arrows.meta,positioning}
\usepackage[hidelinks]{hyperref}
\newtheorem{proposition}{Proposition}

\graphicspath{{figures/}}

\graphicspath{{figures/}}

\begin{document}
\title{DOE-Constrained Distributed Battery Coordination in Unbalanced Distribution Networks}
\author{Bohang Fang, Anderson Toler, 
Lijun Ding, Wenqi Cui, and Cong Chen$^*$
\thanks{Bohang Fang, Anderson Toler, and Cong Chen are with the
Thayer School of Engineering, Dartmouth College,
Hanover, NH, USA.
E-mail: \{Bohang.Fang, Anderson.O.Toler.27, Cong.Chen\}@dartmouth.edu.
Lijun Ding is with the Department of Mathematics,
University of California San Diego,
La Jolla, CA, USA
(e-mail: l2ding@ucsd.edu).
Wenqi Cui is with the Department of Electrical and
Computer Engineering, Tandon School of Engineering,
New York University, Brooklyn, NY, USA
(e-mail: wenqicui@nyu.edu).
\textit{Corresponding author: Cong Chen.}
}%
}

\maketitle

\begin{abstract}
Dynamic operating envelopes (DOEs) provide injection and withdrawal limits for distributed energy resource scheduling while respecting network operational constraints.
This paper develops a risk-aware DOE framework for distributed battery operation in unbalanced three-phase networks using a semidefinite program (SDP). We derive sufficient conditions for exactness of the SDP relaxation, ensuring that the recovered operating points corresponding to the DOE vertices satisfy all voltage and nodal power injection constraints of the original AC network model.
The method computes hourly active-power DOEs under uncertain load conditions and incorporates them directly into a 24-hour battery operation problem.
On a modified IEEE 13-bus feeder, empirical comparisons with unconstrained arbitrage and local Volt--VAR/Volt--Watt control show that DOE-constrained battery scheduling eliminates the observed out-of-sample voltage violations while retaining 94.6\% of the unconstrained arbitrage profit.
\end{abstract}


\section{Introduction}

Distributed residential batteries are increasingly connected to distribution networks and operated by aggregators. Their charging and discharging schedules may cause over- or undervoltage problems, creating a coordination challenge between distribution system operators (DSOs) and battery aggregators \cite{petrou2021integrity}.
Dynamic operating envelopes (DOEs) address this challenge by specifying network injection and withdrawal limits in the real-time operation for distributed energy resources (DERs) \cite{alam2024allocation}. 
In this work, we focus on the application of DOEs to distributed batteries.

Existing DOE methods have limitations in accurately modeling unbalanced three-phase distribution networks while providing rigorous solution guarantees.
Linearized three-phase models \cite{essayeh2024reliable, mahmoodi2024der} are computationally attractive but introduce approximation errors, while convex relaxations of single-phase AC models \cite{moring2026inexactness} cannot capture phase unbalance.
Nonconvex three-phase AC formulations \cite{alam2024allocation} provide accurate network models but generally lack global optimality guarantees.
These limitations become more critical with growing DER penetration, as residential batteries, rooftop photovoltaic systems, and electric vehicle chargers are often connected to individual phases and can further aggravate phase unbalance in distribution networks \cite{wang2020interphase}.

Uncertainty in background demand presents another challenge, since DOEs computed for nominal conditions may not remain feasible under realized operating conditions. Existing studies have addressed such uncertainty using chance-constrained formulations \cite{essayeh2024reliable} and stochastic optimal power flow (OPF) based on polynomial chaos expansion \cite{koirala}. 
However, constructing DOEs that account for such uncertainty while retaining the unbalanced three-phase AC network model and providing operational feasibility guarantees remains computationally challenging and underexplored.

Two practical approaches can be used to impose voltage limits on distributed battery operation. 
One is DOE-constrained scheduling, which sets network-aware limits on battery charging and discharging schedules \cite{petrou2021integrity}. Another is local inverter control under IEEE Standard 1547-2018 \cite{ieee1547}, such as Volt--VAR and Volt--Watt controls, which adjust reactive or active power in response to local voltage measurements.
Such controls can mitigate over- and undervoltage issues while also affecting the economic benefits of battery operation \cite{hashmi2023inverter}. However, the relative voltage performance and economic impacts of DOE-constrained scheduling and local control under common operating conditions remain insufficiently understood, motivating the comparison in this work.

Our main contributions are threefold. 
First, we develop a conditional value-at-risk (CVaR)-based three-phase semidefinite program (SDP) framework for computing active power DOEs under uncertain demand in unbalanced distribution networks, and show that increased phase unbalance can reduce the resulting DOE size.
Second, we establish sufficient conditions for exact SDP relaxation and rank-one recovery at the DOE vertices, ensuring that the recovered operating points satisfy the operational constraints of the original three-phase AC network model at a prescribed reliability level.
Third, we integrate the DOEs into multi-period distributed battery operation and compare DOE-constrained scheduling with unconstrained arbitrage and local Volt--VAR and Volt--Watt controls. Numerical results on a modified IEEE 13-bus feeder show that DOE-constrained scheduling eliminates all observed out-of-sample voltage violations while retaining 94.6\% of the unconstrained arbitrage profit.

\section{Network Model and Feasibility Under Uncertainty}
\label{sec:network_model}
We first introduce the three-phase bus injection model (BIM) and its semidefinite program (SDP) relaxation, referred to as the BIM-SDP model. We then characterize network feasibility for a given realization of the uncertain exogenous injections, which forms the basis for the risk-aware operating envelope construction developed in Section \ref{sec:cvar_doe}.

\subsection{Three-Phase BIM-SDP Model}
\label{subsec:bim}
Consider a radial distribution network represented by an undirected tree
$\mathcal G=(\mathcal N,\mathcal E)$, where
$\mathcal N=\{1,\ldots,N\}$ is the bus set.
Let \(\phi \in \{a,b,c\}\) index the phases.
In this paper, we only consider Y-connected loads.

Neglecting line shunt admittances, let $\mathbf y_{ij}\in\mathbb C^{3\times3}$ denote the series admittance of line $(i,j)\in\mathcal E$, with entries corresponding to absent phases set to zero. The $(i,j)$th $3\times3$ block of the nodal admittance matrix $\mathbf Y\in\mathbb{C}^{3N\times3N}$ is
\begin{equation}
[\mathbf Y]_{ii}=\sum_{k:i\sim k}\mathbf y_{ik},\quad
[\mathbf Y]_{ij}=
\begin{cases}
-\mathbf y_{ij}, & i\sim j,\\
\mathbf 0, & i\nsim j,
\end{cases}\quad i\neq j.
\label{eq:ybus-blocks}
\end{equation}
where $i\sim j$ denotes adjacent buses.

Let $\mathbf V_i:=[V_i^a,V_i^b,V_i^c]^{\mathsf T}\in\mathbb C^3$ denote the complex voltage at bus $i$, and define
$\mathbf V:=[\mathbf V_1^{\mathsf T},\ldots,\mathbf V_N^{\mathsf T}]^{\mathsf T}\in\mathbb C^{3N}$.
Bus 1 is the slack bus with
$\mathbf V_1=\mathbf V_{\rm ref}
:=
[1,\ e^{-\mathrm j2\pi/3},\ e^{\mathrm j2\pi/3}]^{\mathsf T}\ {\rm p.u.}$
Under the BIM formulation, the current injection vector is given by $\mathbf I=\mathbf Y\mathbf V$.
Define the lifted matrix
\begin{equation}
\mathbf W:=\mathbf V\mathbf V^H\in\mathbb H^{3N},
\quad
\mathbf W\succeq0,\quad \operatorname{rank}(\mathbf W)=1.
\label{eq:SDP:W}
\end{equation}
Dropping the rank-one constraint in \eqref{eq:SDP:W} yields the convex
BIM-SDP used in this work.

The complex power injections, squared voltage magnitudes, and active and reactive power injections are given by
\begin{equation}
\begin{aligned}
\mathbf s(\mathbf W)&:=\operatorname{diag}(\mathbf W\mathbf Y^H),
&\qquad
\mathbf v(\mathbf W)&:=\operatorname{diag}(\mathbf W),\\
\mathbf p(\mathbf W)&:=\operatorname{Re}\{\mathbf s(\mathbf W)\},
&
\mathbf q(\mathbf W)&:=\operatorname{Im}\{\mathbf s(\mathbf W)\}.
\end{aligned}
\label{eq:bim-mappings}\nonumber
\end{equation}
where $\mathbf s(\mathbf W)\in\mathbb C^{3N}$ and
$\mathbf p(\mathbf W),\mathbf q(\mathbf W),\mathbf v(\mathbf W)\in\mathbb R^{3N}$ \cite{fang2026scalable}.
The slack bus voltage is fixed through
$[\mathbf W]_{11}=\mathbf M_{\rm ref}:=\mathbf V_{\rm ref}\mathbf V_{\rm ref}^{\sf H}$, where
$[\mathbf W]_{11}\in\mathbb C^{3\times3}$ denotes the top-left block
of $\mathbf W$ associated with bus~1.

\subsection{Network Feasibility Under Uncertain Operating Injections}
\label{subsec:uncertain_feasibility}

Let $\mathbf u\in\mathbb R^{n_u}$ collect the controllable active power injections at the $n_u$ selected bus-phase locations for which the DOE is defined. The lower and upper bounds on $\mathbf u$ define the operating envelope to be computed.
A mapping matrix $\mathbf B\in\mathbb R^{3N\times n_u}$ is used to embed $\mathbf u$ into the full network injection vector, where $\mathbf{B}_{ij}=1$ if the $j$-th controllable injection is located at the $i$-th bus--phase entry and $\mathbf{B}_{ij}=0$ otherwise. 
Let $\omega$ denote an uncertainty realization, and let $\mathbf p^{\rm bg}(\omega)$ and $\mathbf q^{\rm bg}(\omega)$ denote the corresponding uncertain background active and reactive power injections, respectively. 
These background injections may arise from uncertain loads or uncontrollable DERs and are not controlled through the operating envelope.

In addition to $\mathbf p^{\rm bg}(\omega)$, $\mathbf q^{\rm bg}(\omega)$, and $\mathbf u$, let $\mathbf p^{\rm r},\mathbf q^{\rm r}\in\mathbb R^{3N}$ denote recourse active and reactive power injections, representing bounded corrective adjustments available after the uncertainty realization is known.
Their bounds are fixed as
\begin{equation}
\underline{\mathbf p}
\le
\mathbf p^{\rm r}
\le
\overline{\mathbf p},
\qquad
\underline{\mathbf q}
\le
\mathbf q^{\rm r}
\le
\overline{\mathbf q}.
\label{eq:remaining_injection_bounds}
\end{equation}
At bus-phase locations where no recourse is available, the corresponding lower and upper bounds are set to zero.

For a given controllable injection $\mathbf u$ and uncertainty realization $\omega$, the resulting net nodal injections are
\begin{equation}
\begin{aligned}
\mathbf p(\mathbf W)
&=
\mathbf p^{\rm bg}(\omega)
+\mathbf B\mathbf u
+\mathbf p^{\rm r},\\
\mathbf q(\mathbf W)
&=
\mathbf q^{\rm bg}(\omega)
+\mathbf q^{\rm r}.
\end{aligned}
\label{eq:net_injection}
\end{equation}
Eliminating the recourse variables $\mathbf p^{\rm r}$ and $\mathbf q^{\rm r}$ from \eqref{eq:remaining_injection_bounds} and \eqref{eq:net_injection} yields the equivalent nodal power injection constraints
\begin{equation}
\begin{aligned}
\underline{\mathbf p}
&\le
\mathbf p(\mathbf W)
-\mathbf p^{\rm bg}(\omega)
-\mathbf B\mathbf u
\le
\overline{\mathbf p},\\
\underline{\mathbf q}
&\le
\mathbf q(\mathbf W)
-\mathbf q^{\rm bg}(\omega)
\le
\overline{\mathbf q}.
\end{aligned}
\label{eq:power_balance}
\end{equation}

Let $\underline V$ and $\overline V$ denote the lower and upper voltage magnitude limits, respectively. The corresponding nodal power injection and voltage limit residuals are collected in
$\mathbf g(\mathbf W,\mathbf u,\omega)\in\mathbb R^{18N}$ as
\begin{equation}
\mathbf g(\mathbf W,\mathbf u,\omega)
:=
\begin{bmatrix}
\underline{\mathbf p}
-\mathbf p(\mathbf W)
+\mathbf p^{\rm bg}(\omega)
+\mathbf B\mathbf u\\
\mathbf p(\mathbf W)
-\mathbf p^{\rm bg}(\omega)
-\mathbf B\mathbf u
-\overline{\mathbf p}\\
\underline{\mathbf q}
-\mathbf q(\mathbf W)
+\mathbf q^{\rm bg}(\omega)\\
\mathbf q(\mathbf W)
-\mathbf q^{\rm bg}(\omega)
-\overline{\mathbf q}\\
\underline V^2\mathbf 1-\mathbf v(\mathbf W)\\
\mathbf v(\mathbf W)-\overline V^2\mathbf 1
\end{bmatrix},
\label{eq:network_residual}
\end{equation}
where $\mathbf 1\in\mathbb R^{3N}$. Hence,
$\mathbf g(\mathbf W,\mathbf u,\omega)\le\mathbf 0$ enforces the nodal power injection and voltage magnitude limits.

For a given $(\mathbf u,\omega)$, define the SDP-relaxed feasible set as
\begin{equation}
\mathcal X(\mathbf u,\omega)
:=
\left\{
\mathbf W
\;\middle|\;
\begin{array}{l}
\mathbf W\succeq0,\qquad
[\mathbf W]_{11}=\mathbf M_{\rm ref},\\
\mathbf g(\mathbf W,\mathbf u,\omega)\le\mathbf 0
\end{array}
\right\}.
\label{eq:sdp_feasible_set}
\end{equation}
Accordingly, $\mathbf u$ is feasible under the SDP relaxation for realization $\omega$ if
$\mathcal X(\mathbf u,\omega)\neq\varnothing$.

\section{CVaR-Based Operating Envelope Construction}
\label{sec:cvar_doe}

The SDP-relaxed feasibility condition in \eqref{eq:sdp_feasible_set} is defined for a fixed uncertainty realization $\omega$. Requiring this condition to hold for every realization would lead to a robust operating envelope. Instead, we allow limited violations under adverse uncertainty realizations and control their risk through a conditional value-at-risk (CVaR) criterion over sampled realizations. Using this criterion, we construct a box of controllable
injections while controlling the risk of infeasibility under uncertain exogenous injections. 
We consider a single time interval and suppress the time index; in the next section, repeating the calculation with updated load forecasts yields a sequence of dynamic operating envelopes (DOEs).

\subsection{Operating Envelope and Vertex Representation}

The operating envelope at the selected bus-phase locations is
\begin{equation}
\mathcal U
:=
\left\{
\mathbf u\in\mathbb R^{n_u}:
\underline{\mathbf u}\le\mathbf u\le\overline{\mathbf u}
\right\},
\label{eq:doe-box}
\end{equation}
where the inequalities are componentwise and the bounds are decision
variables. For a realization $\omega$, the box is feasible under the SDP relaxation if
\begin{equation}
\mathcal X(\mathbf u,\omega)\neq\varnothing,
\qquad \forall\mathbf u\in\mathcal U.
\label{eq:scenario-doe-feasibility}\nonumber
\end{equation}
Under uncertainty, the desired reliability requirement is 
\begin{equation}
\mathbb P_\omega
\left\{
\mathcal X(\mathbf u,\omega)\neq\varnothing
\ \text{for all }\mathbf u\in\mathcal U
\right\}\ge\alpha,
\label{eq:doe-chance} \nonumber
\end{equation}
where $\alpha\in(0,1)$ is the prescribed reliability level. We measure
the envelope size by the aggregate width
$\mathbf 1^{\mathsf T}
(\overline{\mathbf u}-\underline{\mathbf u})$.

To avoid enforcing network feasibility for every injection in \(\mathcal U\), we exploit the fact that a box is the convex hull of its vertices.
Let
$n_v=2^{n_u}$, $\mathcal K=\{1,\ldots,n_v\}$, and 
let \(\{\mathbf a^k\}_{k\in\mathcal K}\) enumerate the binary vectors.
The $k$-th vertex of the operating envelope is
\begin{equation}
\mathbf u^k
=
\underline{\mathbf u}
+
\mathbf a^k\circ
\left(
\overline{\mathbf u}-\underline{\mathbf u}
\right),
\qquad k\in\mathcal K.
\label{eq:oe_vertices}
\end{equation}
where $\circ$ denotes the elementwise product. For a fixed $\omega$, the constraints in \eqref{eq:sdp_feasible_set} defining $\mathcal X(\mathbf u,\omega)$ are jointly convex in $(\mathbf W,\mathbf u)$, and hence the set of feasible $\mathbf u$ is convex. Since $\mathcal U$ is the convex hull of its vertices, it follows that
\begin{equation}
\mathcal X(\mathbf u,\omega)\neq\varnothing
\quad
\forall\,\mathbf u\in\mathcal U
\quad\Longleftrightarrow\quad
\mathcal X(\mathbf u^k,\omega)\neq\varnothing
\quad
\forall\,k\in\mathcal K.
\label{eq:vertex_feasibility}
\end{equation}
Thus, feasibility of the entire box under the BIM-SDP relaxation can be verified by checking its finitely many vertices.
A distinct feasible network state $\mathbf W^k$ may be associated with each vertex $\mathbf u^k$.

\subsection{CVaR-Based Envelope Optimization}

To control the reliability of the operating envelope under uncertainty, we introduce a CVaR-based formulation. 
Let \(\{\omega_s\}_{s=1}^{n_s}\) denote \(n_s\) sampled uncertainty realizations, with index set \(\mathcal S:=\{1,\ldots,n_s\}\). For each \(s\in\mathcal S\), we introduce a signed feasibility margin \(z_s\in\mathbb R\), common to all vertices of the operating envelope \(\mathcal U\) defined in \eqref{eq:doe-box}, and impose
\[
\mathbf g(\mathbf W^{s,k},\mathbf u^k,\omega_s)\le \mathbf d z_s,
\qquad k\in\mathcal K.
\]

The scaling vector $\mathbf d\in\mathbb R^{18N}$ normalizes operating constraints by their ranges. To avoid zero scaling at bus-phase locations with zero-width injection ranges, we impose strictly positive normalization floors $d_p^{\min}>0$ and $d_q^{\min}>0$.

Hence, $z_s\le0$ ensures satisfaction of the operating limit inequalities in \eqref{eq:network_residual}, whereas $z_s>0$ bounds the normalized constraint violation.
Let \(\mathbf z=[z_1,\ldots,z_{n_s}]^\top\). For equally weighted scenarios, we use the empirical CVaR at reliability level \(\alpha\in(0,1)\) to quantify the upper-tail risk of the scenario-wise feasibility margins over the worst \(1-\alpha\) fraction of scenarios \cite{rockafellar}. We relax the zero-threshold CVaR requirement using a nonnegative allowance \(\tau\) and penalize this allowance in the objective. The operating envelope is then obtained by solving

\begin{subequations}\label{eq:doeproblem}
\begin{align}
\max_{\substack{
\underline{\mathbf u},\,\overline{\mathbf u},\\
\{\mathbf W^{s,k}\},\,\mathbf z,\,\tau
}}
\quad&
\begin{aligned}[t]
&\mathbf1^{\mathsf T}
(\overline{\mathbf u}-\underline{\mathbf u})
-\lambda_\tau\tau\\[-0.3ex]
&\quad-\frac{\beta}{n_sn_v}
\sum_{s\in\mathcal S}\sum_{k\in\mathcal K}
\ell(\mathbf W^{s,k})
\end{aligned}
\label{eq:doe-objective}\\
\mathrm{s.t.}\quad
&\overline{\mathbf u}-\underline{\mathbf u}
\ge\mathbf0,
\label{eq:doe-width}\\
&\mathbf W^{s,k}\succeq0,\qquad
[\mathbf W^{s,k}]_{11}=\mathbf M_{\rm ref},
\label{eq:doe-network}\\
&\mathbf g(\mathbf W^{s,k},\mathbf u^k,\omega_s)
\le
\mathbf d z_s,
\label{eq:doe-scenario-margin}\\
&\operatorname{CVaR}_{\alpha}(\mathbf z)\le\tau,
\qquad \tau\ge0.
\label{eq:doe-cvar-bound}
\end{align}
\end{subequations}
Constraints \eqref{eq:doe-network}--\eqref{eq:doe-scenario-margin} are imposed for all scenarios $s\in\mathcal S$ and vertices $k\in\mathcal K$, with $\mathbf u^k$ defined in \eqref{eq:oe_vertices} and $n_v:=|\mathcal K|$.
The total real power loss is
\begin{equation}
\ell(\mathbf W)
:=
\mathbf 1^{\mathsf T}\mathbf p(\mathbf W)
=
\operatorname{tr}(\mathbf C\mathbf W),
\quad
\mathbf C:=
\frac{1}{2}\left(\mathbf Y+\mathbf Y^{\sf H}\right).
\label{eq:network_loss}\nonumber
\end{equation}

The parameters \(\beta>0\) and \(\lambda_\tau>0\) control the penalties on power loss and the CVaR allowance, respectively.
The choices of $\beta$, $\lambda_\tau$, and $\alpha$ affect the rank-one recovery conditions, as discussed in Appendix~\ref{app:doe-exactness}.

\subsection{DOE Reliability and AC Feasibility}

Following the $\mathcal G$-invertibility framework in \cite{vanderholst2003} and its extension to complex three-phase SDP matrices in \cite{fang2025dispatchable, fang2026scalable}, we analyze the dual of \eqref{eq:doeproblem} and establish sufficient conditions for rank-one recovery at the DOE vertices,
thereby certifying their operational feasibility with respect to the original AC network constraints under scenarios with nonpositive feasibility margins.

\begin{proposition}[Empirical DOE reliability]
\label{prop:doe-exactness}
Suppose \eqref{eq:doeproblem} admits an optimal solution with
$\tau^\star=0$, and let $\mathcal U^\star$ be the resulting
envelope. Under the rank-one recovery conditions in online Appendix~\ref{app:doe-exactness}, for at least an $\alpha$ fraction of the sampled
scenarios, all vertices of $\mathcal U^\star$ are AC feasible
under the original three-phase network model. Moreover, the
entire envelope $\mathcal U^\star$ is feasible under the
BIM--SDP relaxation for those scenarios.
\end{proposition}

The detailed proof is in online Appendix ~\ref{app:doe-exactness}. 
Proposition \ref{prop:doe-exactness} provides theoretical support for the reliability of the three-phase DOE computed in this work, for which no voltage violations are observed in the numerical tests. The proposed formulation also enables efficient DOE computation for real-time operation.

\section{Battery Operation with Local Voltage Control}
\label{sec:battery}

Repeating the calculation of operating envelopes in Section~\ref{sec:cvar_doe} for each time interval yields a sequence of DOEs. These envelopes are incorporated into battery scheduling, which is then compared with local Volt--VAR and Volt--Watt controls following IEEE Standard 1547-2018. Battery capacities are treated as given, and degradation costs are neglected.

\subsection{Battery Arbitrage Scheduling}
\label{subsec:without constraint}

Consider \(n_u\) batteries, one at each bus--phase location represented by the controllable injection vector \(\mathbf u\) in Section~\ref{subsec:uncertain_feasibility}, over the horizon \(\mathcal T:=\{1,\ldots,T\}\). Each interval has duration $\Delta t$. Let $\mathbf p_t^{\rm ch}$ and $\mathbf p_t^{\rm dis}$ denote their charging and discharging powers, and let $\mathbf e_t$ denote the stored energy. The net active power injection is
$\mathbf u_t
=\mathbf p_t^{\rm dis}-\mathbf p_t^{\rm ch},
\ t\in\mathcal T$,
where positive values denote discharging into the network.

For each scheduling interval \(t\in\mathcal T\), let \(\underline{\mathbf u}_t,\overline{\mathbf u}_t\in\mathbb R^{n_u}\) denote the lower and upper DOE bounds, respectively, at the selected battery locations, as computed from \eqref{eq:doeproblem}.

Given buying and selling prices $\lambda_t^{\rm buy}$ and $\lambda_t^{\rm sell}$, the DOE-constrained arbitrage schedule is obtained from
\begin{subequations}\label{eq:batteryproblem}
\begin{align}
\max_{\substack{
\{\mathbf p_t^{\rm ch},\,\mathbf p_t^{\rm dis},\,\mathbf d_t\}_{t\in\mathcal T},\\
\{\mathbf e_t\}_{t=1}^{T+1}
}}
\quad&
\Delta t\sum_{t\in\mathcal T}
\mathbf 1^{\mathsf T}
\left(
\lambda_t^{\rm sell}\mathbf p_t^{\rm dis}
-\lambda_t^{\rm buy}\mathbf p_t^{\rm ch}
\right)
\label{eq:profit}\\[-1.8ex]
\mathrm{s.t.}\quad&
\mathbf e_{t+1}
=\mathbf e_t
+\eta_{\rm ch}\mathbf p_t^{\rm ch}\Delta t
-\frac{\mathbf p_t^{\rm dis}\Delta t}{\eta_{\rm dis}},
\label{eq:soc}\\
&\mathbf 0\le\mathbf p_t^{\rm dis}
\le\overline{\mathbf P}^{\rm dis}\circ\bm \delta_t,
\label{eq:mode-dis}\\
&\mathbf 0\le\mathbf p_t^{\rm ch}
\le\overline{\mathbf P}^{\rm ch}\circ
(\mathbf 1-\bm \delta_t),
\label{eq:mode-ch}\\
&\bm \delta_t\in\{0,1\}^{n_u},
\qquad t\in\mathcal T,
\label{eq:mode}\\
&\underline{\mathbf E}\le\mathbf e_t
\le\overline{\mathbf E},
\quad t=1,\ldots,T+1,
\label{eq:energybounds}\\
&\mathbf e_1=\mathbf E^{\rm ini},
\qquad
\mathbf e_{T+1}=\mathbf E^{\rm target},
\label{eq:terminal}\\
&\underline{\mathbf u}_t
\le
\mathbf p_t^{\rm dis}-\mathbf p_t^{\rm ch}
\le
\overline{\mathbf u}_t,
\qquad t\in\mathcal T.
\label{eq:batterydoe}
\end{align}
\end{subequations}

Here, $\overline{\mathbf P}^{\rm ch}$ and $\overline{\mathbf P}^{\rm dis}$ denote the charging and discharging limits, and $\underline{\mathbf E}$ and $\overline{\mathbf E}$ denote the energy limits.
The binary vector $\bm \delta_t$ prevents simultaneous charging and discharging, making \eqref{eq:batteryproblem} a mixed-integer linear program (MILP). Exact continuous reformulations that avoid binary mode variables have also been studied under suitable conditions \cite{li}; here, we retain the binary formulation for generality.
The DOE bounds enter \eqref{eq:batterydoe} as fixed parameters; omitting \eqref{eq:batterydoe} yields the unconstrained battery arbitrage problem.

\subsection{Local Voltage-Control Benchmarks}
\label{subsec control}

The unconstrained arbitrage schedule,  obtained from \eqref{eq:batteryproblem} by omitting \eqref{eq:batterydoe}, is used as the active power reference for the local Volt--VAR and Volt--Watt controllers.
Local control may curtail the scheduled active power subject to the inverter and battery limits. 
Volt--VAR adjusts reactive power according to the local voltage outside its deadband, with reactive power priority, whereas Volt--Watt operates at zero reactive power and curtails discharging once the voltage exceeds its activation threshold.
Both controllers are evaluated under a quasi-steady-state assumption, with local control settling within each scheduling interval. The operating point is obtained iteratively using nonlinear three-phase power flow, as detailed in online Appendix~\ref{app:local:volt}. The converged active power injections are then used to update the stored energy and realized arbitrage profit.

\section{Numerical Results}

\subsection{Simulation Setup}

The study considers a scaled and modified IEEE 13-bus feeder over a 24-hour horizon.
The voltage magnitude limits are set to $[0.95,1.05]$~p.u.
Two identical batteries are installed at locations 675b and 652a, each with 
an active power rating of 1~kW, an energy capacity of 4~kWh, and an inverter rating of 1.16~kVA. The allowable state-of-charge range is 10\%--90\%, with both the initial and terminal targets set to 30\%, and $\eta_{\rm ch}=\eta_{\rm dis}=95\%$.

At each load bus--phase location, the corresponding entry of active power recourse $\mathbf{p}^r$ is limited to $\pm5\%$ of the hourly forecast demand, with the reactive power recourse bounds set to $0.4$ times the active power recourse bounds. 
The hourly load profile is derived from Pecan Street residential load data \cite{pecanstreet_dataport}. Load uncertainty is modeled using correlated Gaussian multipliers with a pre-truncation standard deviation of 0.08 and pairwise correlation of 0.35, truncated to \([0.75,1.25]\).
Each hourly DOE uses $n_s=1000$ training samples.
Battery arbitrage uses ISO New England day-ahead Hub LMPs on Nov. 4, 2026, with identical buying and selling prices \cite{isone_lmp}.

Four cases are compared. C1 performs unconstrained battery arbitrage, while C2 imposes the proposed DOE on the battery schedule. C3 applies the C1 schedule with local Volt--VAR control, allowing active power curtailment when required by the inverter apparent power limit. C4 applies the C1 schedule with local Volt--Watt control.

\subsection{Results}
\begin{figure}
\centering
\includegraphics[width=1.00 \columnwidth]{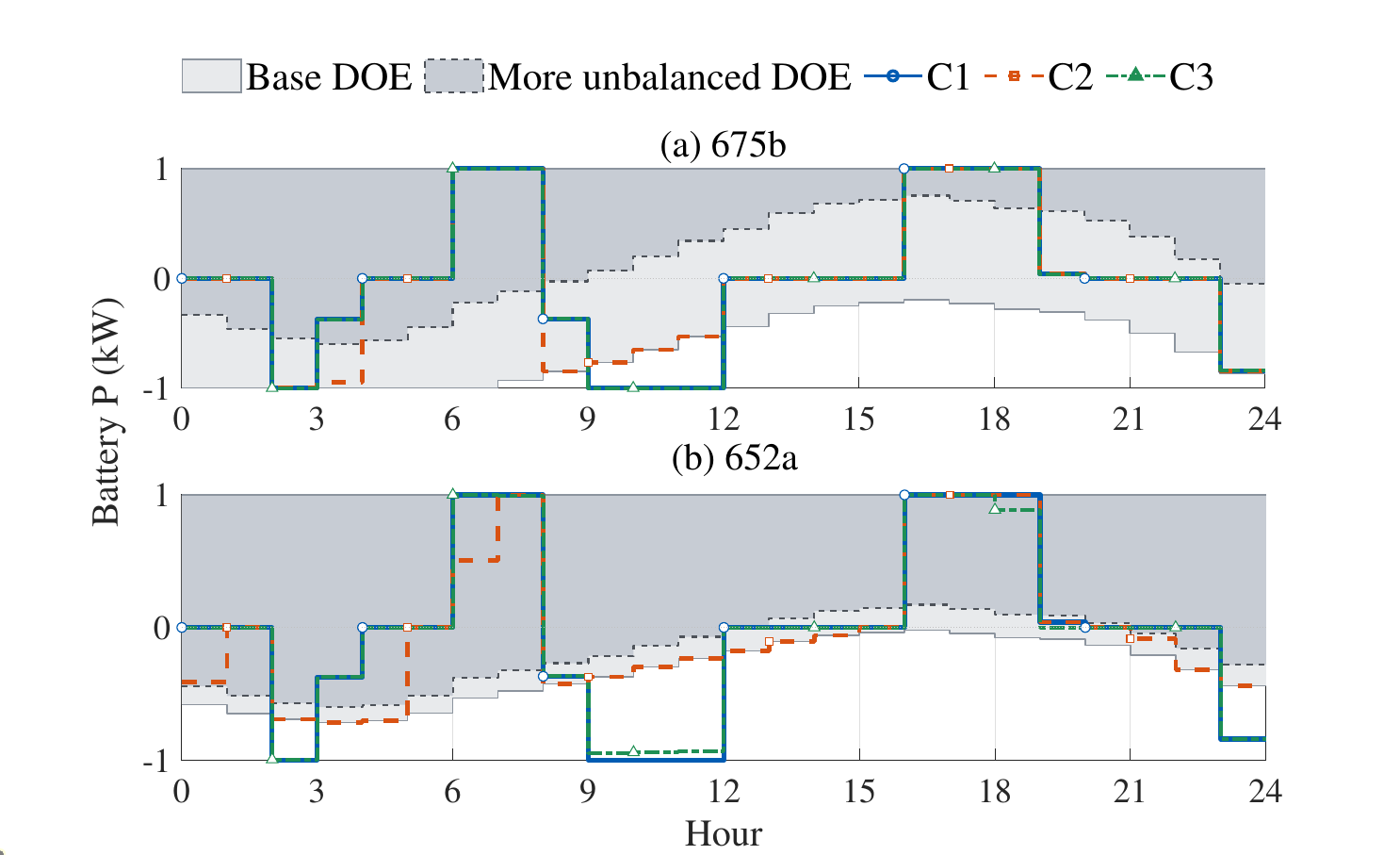}
\caption{Hourly DOE bounds for the base and more unbalanced networks at locations 675b and 652a. The light and dark shaded regions denote the corresponding DOEs, respectively; C1--C3 show the mean executed battery active power for the base network.
}
\label{fig:doe_power}
\end{figure}

We first compare the DOE obtained under the baseline network setting with that under a more unbalanced setting. The latter is constructed by increasing the off-diagonal elements of the admittance matrices and introducing greater imbalance in the phase power injections while keeping the total injection unchanged. As shown in Fig.~\ref{fig:doe_power}, the more unbalanced setting results in a narrower DOE, indicating that increased phase unbalance can reduce the available operating flexibility.
All remaining results are reported for the baseline network setting.

For DOE computation, we set the parameters in \eqref{eq:doe-objective} to $\beta=100$, $\alpha=0.9$, and $\lambda_\tau=1$. 
The resulting solution satisfies $\tau^\star=0$. Moreover, the maximum rank-one ratio $\lambda_2(\mathbf W^{s,k})/\lambda_1(\mathbf W^{s,k})$ over all voltage matrices \(\mathbf W^{s,k}\) across the 24 hourly DOE problems is $5.10\times10^{-9}$, indicating that the SDP solutions are numerically rank-one. 
According to Proposition~\ref{prop:doe-exactness}, such rank-one recovery at the DOE vertices allows AC voltage phasors to be recovered while satisfying the modeled operational constraints for scenarios with nonpositive feasibility margins.

The four operating cases are then evaluated using 5000 test samples for each hour. Fig.~\ref{fig:doe_power} compares the resulting battery active power trajectories with the hourly DOE bounds.
The unconstrained arbitrage schedule C1 exceeds the DOE limits during several hours, whereas C2 modifies the charging and discharging schedule to remain within the calculated envelope. 
Under C3, local Volt--VAR control introduces only limited changes to the active power schedule when reactive power support is constrained by the inverter apparent power rating. 
C4 essentially coincides with C1 because the Volt--Watt controller is not activated: the maximum voltage remains below its activation threshold throughout the simulations. Therefore, C4 is omitted from the figures and tables for clarity.

\begin{figure}
\centering
\includegraphics[width=1.00 \columnwidth]{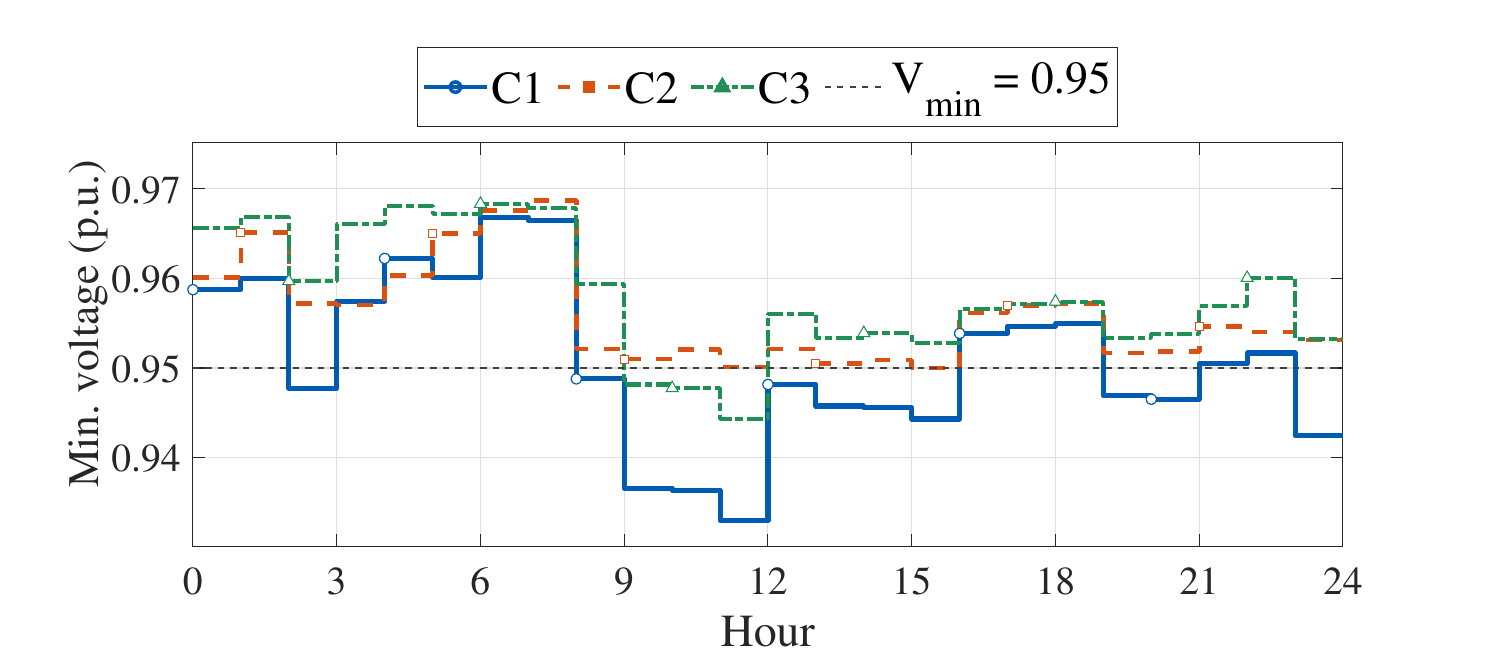}
\caption{Hourly minimum voltage under C1--C3 over all bus locations and 5000 test samples. The dashed line denotes the lower voltage limit of 0.95~p.u.
}
\label{fig:minvolt}
\end{figure}

Fig.~\ref{fig:minvolt} and Table~\ref{tab:performance} summarize the corresponding out-of-sample voltage performance and battery arbitrage profit. C1 achieves the highest profit of \$0.2301/day for the two batteries but results in a voltage violation rate of 9.49\%. Enforcing the proposed DOE in C2 eliminates all observed voltage violations, while reducing the profit to \$0.2176/day, approximately 5.4\% below C1. Local Volt--VAR control in C3 reduces the violation rate to 0.09\% while retaining a profit of \$0.2268/day, but does not completely eliminate voltage limit violations.
The last column of Table~\ref{tab:performance} reports the mean voltage violation depth in p.u., which characterizes the severity of voltage violations in addition to their frequency. Compared with C1, C3 reduces the mean violation depth from $3.7\times10^{-3}$ to $1.2\times10^{-3}$~p.u.

\begin{table}[h]
\centering
\caption{Aggregate battery arbitrage profit and voltage violation under C1--C3.}
\label{tab:performance}
\renewcommand{\arraystretch}{1.2}
\begin{tabular}{c|ccc}
\hline
Case & Profit (\$/day) & Violation (\%) & Mean violation depth (p.u.) \\
\hline
C1 & 0.2301 & 9.49 & $3.7\times10^{-3}$ \\
C2 & 0.2176 & 0      & 0 \\
C3 & 0.2268 & 0.0883  & $1.2\times10^{-3}$ \\
\hline
\end{tabular}
\end{table}

Overall, the proposed DOE eliminates the observed voltage violations at a moderate cost in arbitrage profit. Local Volt--VAR control substantially reduces the violations with a smaller profit loss, but does not eliminate them completely. Volt--Watt control is ineffective in this case because the dominant issue is undervoltage.

\section{Conclusion}

This paper develops a risk-aware DOE framework for distributed battery operation in unbalanced distribution networks using a CVaR-based three-phase SDP formulation. Numerical results show that DOE-constrained scheduling eliminates the observed out-of-sample voltage violations while retaining 94.6\% of the unconstrained arbitrage profit, whereas local Volt--VAR control substantially reduces violations with a smaller economic impact.
The more unbalanced network configuration also yields tighter DOEs, underscoring the importance of three-phase network modeling.
Future work will develop more scalable DOE construction methods for more controllable locations. We will also investigate the convergence of local voltage control and the impact of battery locations on voltage performance and arbitrage profit.

\bibliographystyle{IEEEtran}
\bibliography{references_abbreviated}

\appendices
\section{Proof of Proposition~\ref{prop:doe-exactness}}
\label{app:doe-exactness}

\subsection{Dual Formulation}

Let $c_\ell:=\beta/(n_sn_v)$ and
$c_\alpha:=1/[(1-\alpha)n_s]$.
Associate $\boldsymbol\zeta\ge\mathbf0$ with
$\underline{\mathbf u}-\overline{\mathbf u}\le\mathbf0$,
a free Hermitian multiplier $\boldsymbol\Gamma^{s,k}$ with
$[\mathbf W^{s,k}]_{11}=\mathbf M_{\rm ref}$, and
$\boldsymbol\mu^{s,k}\ge\mathbf0$ with
\eqref{eq:doe-scenario-margin}.
Partition $\boldsymbol\mu^{s,k}$ into the six blocks
$\boldsymbol\mu^{p,-}$, $\boldsymbol\mu^{p,+}$,
$\boldsymbol\mu^{q,-}$, $\boldsymbol\mu^{q,+}$,
$\boldsymbol\mu^{v,-}$, and $\boldsymbol\mu^{v,+}$ in the
order of \eqref{eq:network_residual}, suppressing the indices $(s,k)$
where unambiguous.

The CVaR constraint is equivalently represented by
\begin{equation}
\begin{aligned}
&z_s-\eta-\xi_s\le0,\qquad\xi_s\ge0,\quad s\in\mathcal S,\\
&\eta+c_\alpha\sum_s\xi_s-\tau\le0,\qquad\tau\ge0.
\end{aligned}
\label{eq:app-cvar-epigraph}
\end{equation}
Associate $\theta_s\ge0$ with $z_s-\eta-\xi_s\le0$ and $\rho\ge0$ with $\eta+c_\alpha\sum_s\xi_s-\tau\le0$. The domain constraints $\mathbf W^{s,k}\succeq0$, $\boldsymbol\xi\ge\mathbf0$, and $\tau\ge0$ are retained in the Lagrangian maximization, whereas $\underline{\mathbf u}$, $\overline{\mathbf u}$, $\mathbf z$, and $\eta$ are unrestricted.

For a maximization problem, we subtract each nonnegative
multiplier times its inequality residual. The Lagrangian is
\begin{equation}\nonumber
\begin{aligned}
\mathcal L={}&
\mathbf1^{\mathsf T}
(\overline{\mathbf u}-\underline{\mathbf u})
-\lambda_\tau\tau
-c_\ell\sum_{s,k}\langle\mathbf C,\mathbf W^{s,k}\rangle\\
&-\boldsymbol\zeta^{\mathsf T}
(\underline{\mathbf u}-\overline{\mathbf u})\\
&-\sum_{s,k}
\left\langle\boldsymbol\Gamma^{s,k},
[\mathbf W^{s,k}]_{11}-\mathbf M_{\rm ref}\right\rangle\\
&-\sum_{s,k}(\boldsymbol\mu^{s,k})^{\mathsf T}
\left[\mathbf g(\mathbf W^{s,k},\mathbf u^k,\omega_s)
-\mathbf d z_s\right]\\
&-\sum_s\theta_s(z_s-\eta-\xi_s)
-\rho\left(\eta+c_\alpha\sum_s\xi_s-\tau\right),
\end{aligned}
\label{eq:app-lagrangian}
\end{equation}
where
$\langle\mathbf X,\mathbf Y\rangle
:=\operatorname{Re}\{\operatorname{tr}(\mathbf X^H\mathbf Y)\}$,
and the sums range over $\mathcal S$ and $\mathcal K$.

Let $\mathbf G_m$ denote the Hermitian coefficient matrix of $\mathbf W$ in the $m$-th component of $\mathbf g(\mathbf W,\mathbf u,\omega_s)$. Collecting the terms involving $\mathbf W^{s,k}$ gives
$-\langle\mathbf H^{s,k},\mathbf W^{s,k}\rangle$, where
\begin{equation}
\mathbf H^{s,k}:=
c_\ell\mathbf C
+\operatorname{emb}(\boldsymbol\Gamma^{s,k})
+\sum_m\mu_m^{s,k}\mathbf G_m,
\label{eq:app-H}
\end{equation}
and $\operatorname{emb}(\boldsymbol\Gamma)\in\mathbb H^{3N}$ denotes the embedding of $\boldsymbol\Gamma\in\mathbb H^3$ into the top-left $3\times3$ block, with all remaining entries equal to zero.
Thus the supremum over $\mathbf W^{s,k}\succeq0$ is finite
if and only if $\mathbf H^{s,k}\succeq0$.

To collect the remaining terms, define
\begin{equation}
\begin{aligned}
\mathbf t^{s,k}
&:=\mathbf B^{\mathsf T}
\bigl(\boldsymbol\mu^{p,+}_{s,k}
-\boldsymbol\mu^{p,-}_{s,k}\bigr),\\
\mathbf h_s&:=\mathbf g(\mathbf0,\mathbf0,\omega_s).
\end{aligned}
\label{eq:app-control-coefficient}\nonumber
\end{equation}
Here $\mathbf t^{s,k}$ is the coefficient of $\mathbf u^k$
in $-(\boldsymbol\mu^{s,k})^{\mathsf T}
\mathbf g(\mathbf W^{s,k},\mathbf u^k,\omega_s)$.
Let $\mathbf D^k:=\operatorname{diag}(\mathbf a^k)$, so that
$\mathbf u^k=(\mathbf I-\mathbf D^k)\underline{\mathbf u}
+\mathbf D^k\overline{\mathbf u}$.

Finiteness of the supremum requires the coefficients of the free
variables $\underline{\mathbf u}$, $\overline{\mathbf u}$,
$\mathbf z$, and $\eta$ to vanish, and those of the nonnegative
variables $\boldsymbol\xi$ and $\tau$ to be nonpositive.
The resulting dual is
\begin{equation}
\begin{aligned}
\min_{\substack{
\boldsymbol\zeta,\,\rho,\,\{\theta_s\},\\
\{\boldsymbol\Gamma^{s,k},\,\boldsymbol\mu^{s,k}\}
}}
\quad&
\sum_{s,k}\Bigl[
\langle\boldsymbol\Gamma^{s,k},\mathbf M_{\rm ref}\rangle
-(\boldsymbol\mu^{s,k})^{\mathsf T}\mathbf h_s
\Bigr]\\
\mathrm{s.t.}\quad
&\sum_{s,k}(\mathbf I-\mathbf D^k)\mathbf t^{s,k}
=\mathbf1+\boldsymbol\zeta,\\
&\sum_{s,k}\mathbf D^k\mathbf t^{s,k}
=-\mathbf1-\boldsymbol\zeta,\\
&\theta_s=\sum_k\mathbf d^{\mathsf T}\boldsymbol\mu^{s,k},\\
&\sum_s\theta_s=\rho,\qquad
0\le\rho\le\lambda_\tau,\\
&0\le\theta_s\le c_\alpha\rho,\\
&\mathbf H^{s,k}\succeq0,\qquad
\boldsymbol\mu^{s,k}\ge\mathbf0,\qquad
\boldsymbol\zeta\ge\mathbf0.
\end{aligned}
\label{eq:app-full-dual}
\end{equation}
All indexed constraints hold for every $s\in\mathcal S$
and $k\in\mathcal K$.

Let $d_m$ denote the $m$-th component of the scaling vector
$\mathbf d$. Since $\mathbf d>0$, dual
feasibility gives
\begin{equation}
\begin{aligned}
\kappa_{s,k}:=\mathbf d^{\mathsf T}\boldsymbol\mu^{s,k}
&\le\theta_s\le c_\alpha\lambda_\tau=:\bar\kappa,\\
0\le\mu_m^{s,k}&\le\bar\kappa/d_m.
\end{aligned}
\label{eq:app-multiplier-bound}
\end{equation}
These bounds control the multiplier-dependent term in
\eqref{eq:app-H} and are used in the rank-one recovery
analysis below.

\subsection{Rank-One Recovery and DOE Reliability}

For the analytical recovery result below, consider a full three-phase radial network. 
In this case, $\mathbf H^{s,k}\succeq0$ is
$\mathcal G$-invertible if its off-diagonal blocks associated with network lines are nonsingular, and its blocks between nonadjacent buses vanish \cite{vanderholst2003}.
The latter property follows from the block sparsity of the admittance matrix \(\mathbf Y\) in \eqref{eq:ybus-blocks}.
For each line $(i,j)$, \eqref{eq:app-H} gives
\begin{equation}
\frac{[\mathbf H]_{ij}^{s,k}}{c_\ell}
=
\mathbf C_{ij}
+\sum_m\frac{\mu_m^{s,k}}{c_\ell}[\mathbf G_m]_{ij}.\nonumber
\label{eq:app-edge-bound}
\end{equation}
where \([\cdot]_{ij}\) denotes the \(3\times3\) block corresponding to buses \(i\) and \(j\), i.e., rows \(3i-2\!:\!3i\) and columns \(3j-2\!:\!3j\).

By \eqref{eq:app-multiplier-bound},
\[
\frac{\mu_m^{s,k}}{c_\ell}
\le
\frac{\bar\kappa}{d_m c_\ell}
=
\frac{\lambda_\tau n_v}{(1-\alpha)\beta d_m}.
\]
Hence, since $[\mathbf C]_{ij}$ is nonsingular for every line
$(i,j)$, the same perturbation argument as in
\cite[Prop.~3]{fang2025dispatchable} shows that
$[\mathbf H^{s,k}]_{ij}$ remains nonsingular when
$\lambda_\tau n_v/[(1-\alpha)\beta]$ is sufficiently small.
Therefore, $\mathbf H^{s,k}$ is $\mathcal G$-invertible.

Suppose that the primal--dual pair attains its optima with zero duality gap.
By the KKT complementarity conditions and the reference voltage constraint, we have
\begin{equation}
\mathbf H^{s,k\star}\mathbf W^{s,k\star}=\mathbf0,
\qquad
[\mathbf W^{s,k\star}]_{11}=\mathbf M_{\rm ref}. \nonumber
\label{eq:app-recovery-kkt}
\end{equation}
Under the $\mathcal G$-invertibility condition established above,
the argument in
\cite[Prop.~4 and the proof of Thm.~2]{fang2025dispatchable}
yields $\operatorname{rank}(\mathbf W^{s,k\star})=1$
for every $(s,k)$.
The midpoint argument in \cite[Cor.~1]{fang2026scalable}
also establishes uniqueness of each optimal state matrix.

Next, fix a scenario $s$ with $z_s^\star\le0$ and any
$\mathbf u\in\mathcal U^\star$. Since $\mathcal U^\star$ is the convex hull of its vertices, write
$\mathbf u=\sum_k\vartheta_k\mathbf u^{k\star}$ with
$\vartheta_k\ge0$ and $\sum_k\vartheta_k=1$.
By affinity of the constraints,
$\sum_k\vartheta_k\mathbf W^{s,k\star}$ is SDP feasible for
$\mathbf u$ with the same margin $z_s^\star$.
Hence, $\mathcal U^\star$ is feasible under the BIM--SDP relaxation for every scenario with $z_s^\star\le0$.

Finally, $\tau^\star=0$ implies
$\operatorname{CVaR}_{\alpha}(\mathbf z^\star)\le0$, and thus
\begin{equation}
\frac{1}{n_s}
\left|\{s\in\mathcal S:z_s^\star\le0\}\right|
\ge\alpha. \nonumber
\end{equation}
Together with the vertex AC recovery above, this proves Proposition~\ref{prop:doe-exactness} for the empirical scenario distribution.

\section{Implementation of Local Voltage Control Benchmarks}
\label{app:local:volt}

This appendix details the implementation of the local voltage-control benchmarks in Section~\ref{subsec control}.

\subsection{Volt--VAR Operation with Reactive Power Priority}
\label{subsec:volt_var}

For each scheduling interval $t$, let
$\mathbf u_t^{\rm ref}\in\mathbb R^{n_u}$ denote the scheduled
active power injection computed in \eqref{eq:batteryproblem} and tracked by the local controller, where positive and negative values correspond to discharging and charging, respectively. 
The controller preserves the scheduled charging or discharging direction and permits only reduction in active power magnitude toward zero when required by the battery power and energy limits or the inverter apparent power rating.

Let \(\overline{\mathbf S},\overline{\mathbf Q}\in\mathbb R_+^{n_u}\) denote the inverter apparent power ratings and reactive power limits, respectively, with
\(\mathbf 0\le\overline{\mathbf Q}\le\overline{\mathbf S}\), and define
\begin{equation}
\mathcal Q
:=
\left\{
\mathbf q\in\mathbb R^{n_u}
\;\middle|\;
-\overline{\mathbf Q}
\le\mathbf q
\le\overline{\mathbf Q}
\right\}. \nonumber
\label{eq:q_feasible_set}
\end{equation}

The feasible active power set is constructed as the intersection of three component sets.
The reference consistent set is
\begin{equation}
\mathcal U_t^{\rm ref}
:=
\left\{
\mathbf u
\;\middle|\;
\min\{\mathbf 0,\mathbf u_t^{\rm ref}\}
\le\mathbf u
\le
\max\{\mathbf 0,\mathbf u_t^{\rm ref}\}
\right\}, \nonumber
\label{eq:reference_consistent_set}
\end{equation}
which preserves the scheduled charging or discharging direction and
prevents the controller from increasing its magnitude. 
At the beginning of interval $t$, the actual stored energy
$\mathbf e_t^{\rm act}$ is known and remains fixed throughout the
inner control iteration.
The battery
feasible set is
\begin{equation}
\mathcal U_t^{\rm bat}
:=
\left\{
\mathbf u
\;\middle|\;
\begin{array}{l}
-\overline{\mathbf P}^{\rm ch}
\le\mathbf u
\le\overline{\mathbf P}^{\rm dis},
\\[0.6ex]
-\dfrac{\overline{\mathbf E}-\mathbf e_t^{\rm act}}
{\eta_{\rm ch}\Delta t}
\le\mathbf u
\le
\dfrac{\eta_{\rm dis}
(\mathbf e_t^{\rm act}-\underline{\mathbf E})}
{\Delta t}
\end{array}
\right\}, \nonumber
\label{eq:battery_feasible_set}
\end{equation}
which enforces the battery power and energy limits. For a given
$\mathbf q\in\mathcal Q$, the inverter feasible active power set is
\begin{equation}
\mathcal U^{\rm inv}(\mathbf q)
:=
\left\{
\mathbf u
\;\middle|\;
\mathbf u\circ\mathbf u
+
\mathbf q\circ\mathbf q
\le
\overline{\mathbf S}\circ\overline{\mathbf S}
\right\}.
\label{eq:inverter_feasible_set} \nonumber
\end{equation}
The admissible active power set is therefore
\begin{equation}
\mathcal I_t(\mathbf q)
:=
\mathcal U_t^{\rm ref}
\cap
\mathcal U_t^{\rm bat}
\cap
\mathcal U^{\rm inv}(\mathbf q).
\label{eq:active_feasible_set} \nonumber
\end{equation}
All vector inequalities are componentwise. During the inner
iteration, $\mathbf u_t^{\rm ref}$ and $\mathbf e_t^{\rm act}$ remain
fixed, while $\mathcal I_t(\mathbf q)$ changes with the updated
reactive power injection.
Within interval \(t\), let \(k\) index the quasi-steady-state inner control iteration. At iteration \(k\), the nonlinear three-phase power flow is solved using the current battery injections \((\mathbf u_t^k,\mathbf q_t^k)\). The resulting battery terminal voltage magnitudes are passed through the prescribed Volt--VAR characteristic to obtain the reactive power command \(\mathbf q_t^{\mathrm{cmd},k}\) \cite{narang}. Throughout the inner iteration, \(\mathbf u_t^{\rm ref}\) and \(\mathbf e_t^{\rm act}\) remain fixed, whereas \(\mathcal I_t(\mathbf q)\) changes with the updated reactive power injection.

Using a relaxed update, the battery operating point is updated as
\begin{subequations}\label{eq:fixedpoint}
\nonumber
\begin{align}
\mathbf q_t^{k+1}
&=
\operatorname{proj}_{\mathcal Q}
\left[
\rho\mathbf q_t^k
+
(1-\rho)\mathbf q_t^{{\rm cmd},k}
\right],
\label{eq:q_update}
\\
\mathcal I_t^{k+1}
&:=
\mathcal I_t(\mathbf q_t^{k+1}),
\label{eq:set_update}
\\
\mathbf u_t^{{\rm tar},k+1}
&=
\operatorname{proj}_{\mathcal I_t^{k+1}}
\left(\mathbf u_t^{\rm ref}\right),
\label{eq:active_target}
\\
\mathbf u_t^{k+1}
&=
\operatorname{proj}_{\mathcal I_t^{k+1}}
\left[
\rho\mathbf u_t^k
+
(1-\rho)\mathbf u_t^{{\rm tar},k+1}
\right].
\label{eq:u_update}
\end{align}
\end{subequations}
Here, $\rho\in[0,1)$ is a relaxation factor. A larger $\rho$ gives
smaller updates and can mitigate oscillations, whereas $\rho=0$
corresponds to an undamped update.

The reactive power injection is updated first, after which the
updated $\mathbf q_t^{k+1}$ determines 
the inverter apparent power headroom available for active power and hence the new feasible set
$\mathcal I_t^{k+1}$. The reference schedule is projected onto this
set to obtain $\mathbf u_t^{{\rm tar},k+1}$, and the active power
iterate is moved toward this target while remaining in the updated
set. The resulting $(\mathbf u_t^{k+1},\mathbf q_t^{k+1})$ is then
used in the next power flow evaluation, closing the Volt--VAR control
loop.

\subsection{Volt--Watt Operation}
\label{subsec:volt_watt}

Volt--Watt operation maintains zero reactive power,
$\mathbf q_t=\mathbf0$. At iteration $k$, the nonlinear three-phase
power flow is solved using the current active power injection
$\mathbf u_t^k$. Let $\mathbf v_t^k$ denote the resulting battery
terminal voltage magnitudes. The Volt--Watt
characteristic maps $\mathbf v_t^k$ to the discharging power limits
$\overline{\mathbf p}_t^{{\rm VW},k}$ \cite{narang}, which decrease
at high voltage. Charging is not modified by the Volt--Watt
characteristic.

Define the Volt--Watt feasible set as
\begin{equation}
\mathcal U_t^{{\rm VW},k}
:=
\left\{
\mathbf u
\;\middle|\;
\mathbf u\le
\overline{\mathbf p}_t^{{\rm VW},k}
\right\},
\label{eq:volt_watt_set}\nonumber
\end{equation}
and the resulting active power feasible set as
\begin{equation}
\mathcal I_t^{{\rm VW},k}
:=
\mathcal I_t(\mathbf0)
\cap
\mathcal U_t^{{\rm VW},k}.
\label{eq:volt_watt_interval}\nonumber
\end{equation}
Thus, the battery and inverter limits introduced in
Section~\ref{subsec:volt_var} remain enforced, while the additional
Volt--Watt constraint limits discharging according to the local
voltage.

The active power update is
\begin{subequations}\label{eq:volt_watt_update}\nonumber
\begin{align}
\mathbf u_t^{{\rm tar},k+1}
&=
\operatorname{proj}_{\mathcal I_t^{{\rm VW},k}}
\left(\mathbf u_t^{\rm ref}\right),
\label{eq:vw_target}
\\
\mathbf u_t^{k+1}
&=
\operatorname{proj}_{\mathcal I_t^{{\rm VW},k}}
\left[
\rho\mathbf u_t^k
+
(1-\rho)\mathbf u_t^{{\rm tar},k+1}
\right].
\label{eq:vw_update}
\end{align}
\end{subequations}
Because $\overline{\mathbf p}_t^{{\rm VW},k}$ depends on the voltage
obtained from the current power flow, the feasible set is updated at
every iteration. The resulting $\mathbf u_t^{k+1}$ is then used in
the next power flow evaluation, closing the Volt--Watt control loop.
The same relaxation factor and convergence criteria as in the
Volt--VAR controller are used.

\paragraph{Realized battery operation and cash flow}
For either local controller, let $\mathbf u_t^{\rm act}$ denote the
final active power injection. The realized discharging and charging
powers are
\begin{equation}
\mathbf p_t^{\rm dis,act}
=
[\mathbf u_t^{\rm act}]_+,
\qquad
\mathbf p_t^{\rm ch,act}
=
[-\mathbf u_t^{\rm act}]_+, \nonumber
\label{eq:realized_battery_powers}
\end{equation}
where the positive part is taken componentwise.
Starting from
$\mathbf e_1^{\rm act}=\mathbf E^{\rm ini}$, the actual stored energy
is updated using \eqref{eq:soc} with these realized powers, and the
corresponding cashflow is evaluated using \eqref{eq:profit}.
No separate remuneration is assigned to reactive power.
Since local control may curtail the scheduled active power,
the realized terminal energy may differ from
$\mathbf E^{\rm target}$; this deviation is reported together with
the realized cashflow.

\end{document}